\documentclass[aps,prl,reprint,amsmath,amssymb,showpacs,nobibnotes]{revtex4-1}
\usepackage{graphicx}
\usepackage{hyperref} 
\usepackage{dsfont}    
\usepackage{color}

\newcommand\trick[1]{}

\newcommand{\be}{\begin{equation}}
\newcommand{\ee}{\end{equation}}

\newcommand{\bit}{\begin{itemize}}  \newcommand{\eit}{\end{itemize}}
\newcommand{\ben}{\begin{enumerate}}  \newcommand{\een}{\end{enumerate}}

\newcommand{\rf}[1]{(\ref{#1})}

\def\bd{\begin{document}}
\def\ed{\end{document}}
\def\bea{\begin{eqnarray}}
\def\eea{\end{eqnarray}}
\let\bm=\bibitem

\def\la{\langle}
\def\ra{\rangle}

\def\npb#1#2#3{Nucl. Phys. {\bf{B#1}} #3 (#2)}
\def\plb#1#2#3{Phys. Lett. {\bf{#1B}} #3 (#2)}
\def\prl#1#2#3{Phys. Rev. Lett. {\bf{#1}} #3 (#2)}
\def\prd#1#2#3{Phys. Rev. {D bf{#1}} #3 (#2)}
\def\cmp#1#2#3{Comm. Math. Phys. {\bf{#1}} #3 (#2)}
\def\cqg#1#2#3{Class. Quantum Grav. {\bf{#1}} #3 (#2)}
\def\nppsa#1#2#3{Nucl. Phys. B (Proc. Suppl.) {\bf{#1A}}#3 (#2)}
\def\ap#1#2#3{Ann. of Phys. {\bf{#1}} #3 (#2)}
\def\ijmp#1#2#3{Int. J. Mod. Phys. {\bf{A#1}} #3 (#2)}
\def\rmp#1#2#3{Rev. Mod. Phys. {\bf{#1}} #3 (#2)}
\def\mpla#1#2#3{Mod. Phys. Lett. {\bf A#1} #3 (#2)}
\def\jhep#1#2#3{J. High Energy Phys. {\bf #1} #3 (#2)}
\def\atmp#1#2#3{Adv. Theor. Math. Phys. {\bf #1} #3 (#2)}

\def\N{{\cal N}}
\def\sst{\scriptscriptstyle}
\def\thetabar{\bar\theta}
\def\Tr{{\rm Tr}}
\def\one{\mbox{1 \kern-.59em {\rm l}}}

\def\a{\alpha}      \def\da{{\dot\alpha}}  \def\dA{{\dot A}}
\def\b{\beta}       \def\db{{\dot\beta}}
\def\g{\gamma}  \def\G{\Gamma}  \def\dc{{\dot\gamma}}
\def\d{\delta}  \def\D{\Delta}  \def\ddt{\dot\delta}
\def\e{\epsilon}
\def\ve{\varepsilon}
\def\uve{\upvarepsilon}
\def\f{\phi}    \def\F{\Phi}    \def\vvf{\f}
\def\vphi{\varphi}
\def\h{\eta}
\def\k{\kappa}
\def\l{\lambda} \def\L{\Lambda}
\def\m{\mu} \def\n{\nu}
\def\o{\omega}
\def\p{\pi} \def\P{\Pi}
\def\r{\rho}
\def\s{\sigma}  \def\S{\Sigma}
\def\t{\tau}
\def\th{\theta} \def\Th{\Theta} \def\vth{\vartheta}
\def\X{\Xeta}
\def\z{\zeta}

\def\na{\nabla}

\def\cA{{\cal A}} \def\cB{{\cal B}} \def\cC{{\cal C}}
\def\cD{{\cal D}} \def\cE{{\cal E}} \def\cF{{\cal F}}
\def\cG{{\cal G}} \def\cH{{\cal H}} \def\cI{{\cal I}}
\def\cJ{{\mathscr J}} \def\cK{{\cal K}} \def\cL{{\cal L}}
\def\cM{{\cal M}} \def\cN{{\cal N}} \def\cO{{\cal O}}
\def\cP{{\cal P}} \def\cQ{{\cal Q}} \def\cR{{\cal R}}
\def\cS{{\cal S}} \def\cT{{\cal T}} \def\cU{{\cal U}}
\def\cV{{\cal V}} \def\cW{{\cal W}} \def\cX{{\cal X}}
\def\cY{{\cal Y}} \def\cZ{{\cal Z}}

\def\ua{\underline{\alpha}}
\def\uc{\underline{\phantom{\alpha}}\!\!\!\gamma}
\def\um{\underline{\mu}}
\def\ud{\underline\delta}
\def\ue{\underline\epsilon}
\def\una{\underline a}\def\unA{\underline A}
\def\unb{\underline b}\def\unB{\underline B}
\def\unc{\underline c}\def\unC{\underline C}
\def\und{\underline d}\def\unD{\underline D}
\def\une{\underline e}\def\unE{\underline E}
\def\unf{\underline{\phantom{e}}\!\!\!\! f}\def\unF{\underline F}
\def\unm{\underline m}\def\unM{{\underline M}}
\def\unn{\underline n}\def\unN{{\underline N}}
\def\unp{\underline{\phantom{a}}\!\!\! p}\def\unP{\underline P}
\def\unq{\underline{\phantom{a}}\!\!\! q}
\def\unQ{\underline{\phantom{A}}\!\!\!\! Q}
\def\unH{\underline{H}}

\def\As {{A \hspace{-6.4pt} \slash}\;}
\def\bs {{b \hspace{-6.4pt} \slash}\;}
\def\Ds {{D \hspace{-6.4pt} \slash}\;}
\def\Gts {{\Gt \hspace{-6.4pt} \slash}\;}
\def\ds {{\del \hspace{-6.4pt} \slash}\;}
\def\ss {{\s \hspace{-6.4pt} \slash}\;}
\def\ks {{ k \hspace{-6.4pt} \slash}\;}
\def\ps {{p \hspace{-6.4pt} \slash}\;}
\def\xs {{x \hspace{-6.4pt} \slash}\;}
\def\pas {{{p_1} \hspace{-6.4pt} \slash}\;}
\def\pbs {{{p_2} \hspace{-6.4pt} \slash}\;}
\def\cFs {{{\cal F} \hspace{-6.4pt} \slash}\;}
\def\Dss {{D \hspace{-7.5pt} \slash}\;}
\def\dss {{\del \hspace{-7.0pt} \slash}\;}

\def\Ah{{\hat{A}}}
\def\Dh{{\hat{D}}}
\def\Gh{{\hat{G}}}
\def\Fh{{\hat{F}}}
\def\Ih{{\hat{I}}}
\def\Jh{{\hat{J}}}
\def\Kh{{\hat{K}}}
\def\Lh{{\hat{L}}}
\def\Ph{{\hat{P}}}
\def\Rh{{\hat{R}}}
\def\Vh{{\hat{V}}}
\def\Xh{{\hat{X}}}

\def\ah{{\hat{\a}}}
\def\bh{{\hat{\b}}}
\def\gh{{\hat{\g}}}
\def\dh{{\hat{\d}}}
\def\rh{{\hat{\r}}}
\def\hh{\hat{h}}
\def\uh{\hat{u}}
\def\xh{\hat{x}}
\def\yh{\hat{y}}
\def\ph{\hat{p}}
\def\xih{\hat{\xi}}
\def\chih{\hat{\chi}}
\def\Psih{\hat{\Psi}}
\def\phih{\hat{\phi}}

\def\psit{\tilde{\psi}}
\def\Psit{\tilde{\Psi}}
\def\Psibt{\tilde{\bar{Psi}}}

\def\lambdat{\tilde {\lambda}}
\def\st{\tilde{\sigma}}

\def\delt{\tilde{\delta}}
\def\Phit{\tilde{\Phi}}
\def\Phitb{\overline{\tilde{Phi}}}
\def\tht{\tilde{\th}}
\def\lt{\tilde{\l}}
\def\chit{\tilde{\chi}}
\def\phit{\tilde{\phi}}

\def\At{\tilde{A}}
\def\Bt{\tilde{B}}
\def\Ct{\tilde{C}}
\def\Dt{\tilde{D}}
\def\Et{\tilde{E}}
\def\Ft{\tilde{F}}
\def\Gt{\tilde{G}}
\def\Ht{\tilde{H}}
\def\It{\tilde{I}}
\def\Jt{\tilde{J}}
\def\Qt{\tilde{Q}}
\def\Rt{\tilde{R}}
\def\Mt{\tilde{M }}
\def\Nt{\tilde{N}}
\def\St{\tilde{S}}
\def\Vt{\tilde{V}}
\def\Xt{\tilde{X}}
\def\at{\tilde{a}}
\def\ct{\tilde{c}}
\def\dt{\tilde{d}}
\def\htt{\tilde{h}}
\def\ft{\tilde{f}}
\def\gt{\tilde{g}}
\def\pt{\tilde{p}}
\def\qt{\tilde{q}}
\def\vt{\tilde{v}}
\def\nt{\tilde{n}}
\def\ut{\tilde{u}}
\def\wt{\tilde{w}}
\def\zt{\tilde{z}}
\def\xt{\tilde{x}}
\def\yt{\tilde{y}}
\def\Psit{\tilde{\Psi}}
\def\vphit{\tilde{\varphi}}
\def\tD{\tilde{\D}}

\def\eb{\bar{\epsilon}}
\def\delb{\bar{\partial}}
\def\thb{\bar{\theta}}
\def\mub{\bar{\mu}}
\def\lamb{\bar{\l}}
\def\psib{\bar{\psi}}
\def\sb{\bar{\sigma}}
\def\xib{\bar{\xi}}
\def\chib{\bar{\chi}}

\def\Psib{\bar{\Psi}}
\def\Phib{\bar{\Phi}}
\def\Lamb{\bar{\Lambda}}
\def\Sb{{\overline \Sigma}}
\def\cb{\bar{c}}
\def\hb{\bar{h}}
\def\qb{\bar{q}}
\def\wb{\bar{w}}
\def\ub{\bar{u}}
\def\zb{{\bar{z}}}
\def\Hb{\bar{H}}
\def\Qb{{\bar Q}}
\def\Omegab{\overline{\Omega}}
\def\ob{\overline{\omega}}

\def\Ab{{\overline A}} \def\Bb{{\overline B}} \def\Cb{{\overline C}}
\def\Db{{\overline D}} \def\Eb{{\overline E}} \def\Fb{{\overline F}}
\def\Gb{{\overline G}}
\def\Ib{{\overline I}}
\def\Jb{{\overline J}} \def\Kb{{\overline K}} \def\Lb{{\overline L}}
\def\Mb{{\overline M}} \def\Nb{{\overline N}} \def\Ob{{\overline O}}
\def\Pb{{\overline P}}  \def\Rb{{\overline R}}
 \def\Tb{{\overline T}} \def\Ub{{\overline U}}
\def\Vb{{\overline V}} \def\Wb{{\overline W}} \def\Xb{{\overline X}}
\def\Yb{{\overline Y}} \def\Zb{{\overline Z}}

\def\fb{{\overline f}}
\def\gb{{\overline g}}
\def\mb{{\overline m}}
\def\lb{{\overline l}}
\def\yb{{\overline y}}

\def\ldel{{\overleftarrow{\del}}}
\def\rdel{{\overrightarrow{\del}}}
\def\ldeldel{{\overleftarrow{\del^2}}}
\def\rdeldel{{\overrightarrow{\del^2}}}
\def\ldelb{{\overleftarrow{\bar{\del}}}}
\def\rdelb{{\overrightarrow{\bar{\del}}}}

\def\ba{{\bf a}}
\def\bk{{\bf k}}
\def\bl{{\bf l}}
\def\bp{{\bf p}}
\def\bq{{\bf q}}
\def\br{{\bf r}}
\def\bt{{\bf t}}
\def\bu{{\bf u}}
\def\bv{{\bf v}}
\def\bx{{\bf x}}
\def\by{{\bf y}}
\def\bA{{\bf A}}
\def\bR{{\bf R}}
\def\bV{{\bf V}}

\def\bz{{\boldsymbol{\zeta}}}

\def\bone{{\bf 1}}

\def\va{{\vec a}}
\def\vk{{\vec k}}
\def\vp{{\vec p}}
\def\vq{{\vec q}}
\def\vx{{\vec x}}
\def\vy{{\vec y}}
\def\vu{{\vec u}}
\def\vv{{\vec v}}
\def \vH{{\vec H}}
\def \vg{{\vec g}}

\def\vs{{\vec \sigma}}
\def\vtau{{\vec \tau}}

\newcommand{\ov}[1]{\overrightarrow{#1}}

\def\frA{\mathfrak{A}}
\def\frB{\mathfrak{B}}
\def\frC{\mathfrak{C}}
\def\frD{\mathfrak{D}}
\def\frE{\mathfrak{E}}
\def\frF{\mathfrak{F}}
\def\frG{\mathfrak{G}}
\def\frH{\mathfrak{H}}
\def\frM{\mathfrak{M}}
\def\frN{\mathfrak{N}}
\def\frR{\mathfrak{R}}
\def\frW{\mathfrak{W}}

\def\fra{\mathfrak{a}}
\def\frb{\mathfrak{b}}
\def\frf{\mathfrak{f}}
\def\frg{\mathfrak{g}}
\def\frh{\mathfrak{h}}
\def\frl{\mathfrak{l}}
\def\frs{\mathfrak{s}}
\def\fri{\mathfrak{i}}
\def\frj{\mathfrak{j}}

\def\ma{\mathfrak{a}}
\def\mg{\mathfrak{g}}
\def\mh{\mathfrak{h}}
\def\mR{\mathfrak{R}}
\def\mN{\mathfrak{N}}

\newcommand{\nn}{{\nonumber}}

\def\d{\delta}\def\D{\Delta}\def\ddt{\dot\delta}

\def\pa{\partial} \def\del{\partial}
\def\xx{\times}
\def\uno{\mbox{1 \kern-.59em {\rm l}}}

\def\trp{^{\top}}
\def\inv{^{-1}}
\def\dag{\dagger}
\def\pr{^{\prime}}

\def\rar{\rightarrow}
\def\lar{\leftarrow}
\def\lrar{\leftrightarrow}

\newcommand{\0}{\,\!}      
\def\one{1\!\!1\,\,}
\def\im{\imath}
\def\jm{\jmath}

\newcommand{\tr}{\mbox{tr}}
\newcommand{\slsh}[1]{/ \!\!\!\! #1}

\newcommand{\1}{\mbox{1}\hspace{-0.25em}\mbox{l}}

\def\vac{|0\rangle}
\def\lvac{\langle 0|}

\def\hlf{\frac{1}{2}}
\def\ove#1{\frac{1}{#1}}
\newcommand{\hot}[1]{\frac{#1}{2}}

\def\Box{\square}
\def\CC {\mathbb{C}}
\def\FF {\mathbb{F}}
\def\RR{\mathbb{R}}
\def\NN{\mathbb{N}}
\def\ZZ{\mathbb{Z}}
\def\bb#1{{\bf #1}}
\def\bcomment#1{}
\def\bfhat#1{{\bf \hat{#1}}}
\def\VEV#1{\left\langle #1\right\rangle}

\newcommand{\ex}[1]{{\rm e}^{#1}} \def\ii{{\rm i}}

\newcommand{\lrbrk}[1]{\left(#1\right)}
\newcommand{\lrsbrk}[1]{\left[#1\right]}
\newcommand{\sfrac}[2]{{\textstyle\frac{#1}{#2}}}

\def\stw{{\sqrt{2}}}

\def\rf {{\rm f}}
\def\ri {{\rm i}}
\def\rj {{\rm j}}
\def\rn {{\rm n}}
\def\rk {{\rm k}}
\def\rl {{\rm l}}
\def\rr {{\rm r}}
\def\rs {{\scriptscriptstyle \rm S}}
\def\rt {{\scriptscriptstyle \rm T}}

\def\rQ {{\scriptscriptstyle \rm \cQ}}
\def\rR {{\scriptscriptstyle \rm \cR}}

\def\cQb{{\cal \Qb}}
\def\cRb{{\cal \Rb}}
\def\cWb{{\cal \Wb}}

\def\fd {{\rm N}}
\def\afd {{\overline{\rm N}}}

\def \II {I\hspace{-.1em}I\hspace{.1em}}
\def \IIA {\mbox{\II A\hspace{.2em}}}
\def \IIB {\mbox{\II B\hspace{.2em}}}
\def \gs {g^s}
\def \ls {\lambda^s}

\def \I {{\cal I}}
\def \qs {q\hspace{-.53em}/\hspace{.15em}}
\def \ks {k\hspace{-.53em}/\hspace{.15em}}
\def \YM {{\mbox{\tiny YM}}}
\def \gym {g_{\YM}}

\def \Lc {\L_c}
\def\IR{\relax{\rm I\kern-.18em R}}
\def \id {{\bf 1}}

\def\cci{\ell}
\def\ccj{\ell'}

\def\bbq{\pmb{q}}
\def\bom{\pmb{\o}}
\def\bJ{\pmb{J}}
\def\bM{\pmb{M}}
\def\bB{\pmb{B}}
\def\bn{\pmb{n}}
\def\bE{\pmb{E}}

\newcommand{\rrr}[1]{\vskip 0.2cm \noindent{\it #1} ---}

\begin{document}

\title{Entanglement Island versus Massless Gravity}
\author{Rong-Xin Miao}
\affiliation{School of Physics and Astronomy, Sun Yat-Sen University, Zhuhai, 519082, China}


\begin{abstract}
Entanglement islands play an essential role in the recent breakthrough in addressing the black hole information paradox. Inspired by double holography, it is conjectured that the entanglement islands can exist only in massive gravity. There are many pieces of evidence but also debates for this conjecture. This paper recovers the massless entanglement island in wedge holography with negative DGP gravity on the brane. However, the spectrum of negative DGP gravity includes a massive ghost, implying the model is unstable. Our work supports the view that there is no entanglement island in a well-defined braneworld model of massless gravity if one divides the radiation and black hole regions by minimizing entanglement entropy. However, such a partition results in a zero radiation region containing no information. Whether there are other physical non-trivial partitions of the radiation region is an open question and deserves further study.
\end{abstract}

\maketitle

\rrr{Introduction}
The black hole information problem \cite{Hawking:1976ra} presents a sharp contradiction between general relativity and quantum mechanics, whose resolution may open a window for their unification. Recently, a significant breakthrough in solving this problem has been achieved, where the entanglement island plays an essential role \cite{Penington:2019npb, Almheiri:2019psf, Almheiri:2019hni}. See \cite{Almheiri:2020cfm} for a review on this topic. So far, the discussions focus on either Jackiw–Teitelboim gravity in 1+1 dimensions without gravitons or double holography \cite{Karch:2000ct, Takayanagi:2011zk} in higher dimensions with only massive gravitons on the branes \cite{Almheiri:2019psy, Geng:2020qvw}. In one novel doubly holographic model called wedge holography \cite{Akal:2020wfl}, there is massless gravity on the brane \cite{Hu:2022lxl} but no entanglement island \cite{Geng:2020fxl, Geng:2021hlu, Geng:2022fui,Geng:2023qwm,Geng:2023iqd}. Inspired by the above facts, \cite{ Geng:2021hlu, Geng:2022fui} conjectures that the entanglement island can exist only in massive gravity theories. They give a proof that the island is inconsistent with gravitational Gauss's law for long-range gravity \cite{Geng:2021hlu}. There are debates on this problem; see \cite{Krishnan:2020fer, Ghosh:2021axl} for example. See also the general physical picture of islands \cite{Almheiri:2020cfm}, which works for long-range gravity naturally.

It is crucial whether the island mechanism can apply to massless gravity. First, gravity in the real world is massless. Gravitational wave experiments impose a strict upper bound on the gravity mass \cite{LIGOScientific:2016lio}. Second, massive gravity suffers severe theoretical problems. Although it can be made ghost-free \cite{deRham:2010kj}, it includes superluminal solutions and thus is non-causal \cite{Deser:2012qx}. Therefore the significance of recent breakthroughs is discounted if it works only for massive gravity. 
This paper shows that the massless island can exist in wedge holography with negative Dvali-Gabadadze-Porrati (DGP) gravity \cite{Dvali:2000hr}.  However, negative DGP gravity yields a massive ghost on the brane \cite{Miao:2023mui}, which is not a well-defined theory. 

\begin{figure}[t]
\centering
\includegraphics[width=7cm]{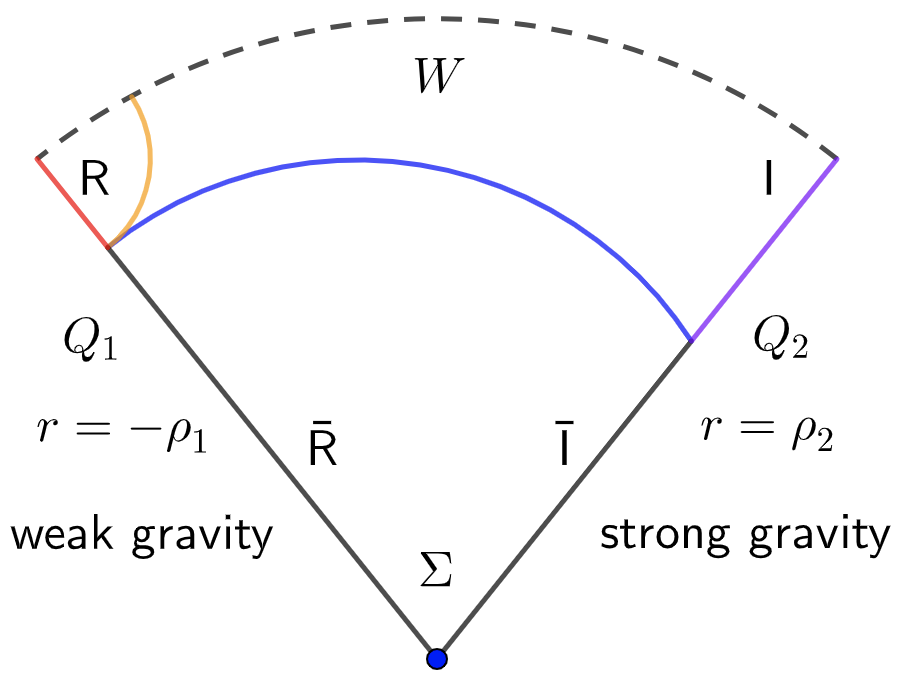}
\caption{Geometry of wedge holography and its interpretation in black hole information paradox. $Q_1$ denotes the bath brane with weak gravity, and $Q_2$ is the AdS brane with intense gravity. The red and purple lines denote the radiation R and island I on branes. The dotted line, blue, and orange lines indicate the horizon, RT surfaces in the island and no-island phase at $t=0$, respectively. }
\label{Wedge}
\end{figure}
\begin{figure}[t]
\centering
\includegraphics[width=7.6cm]{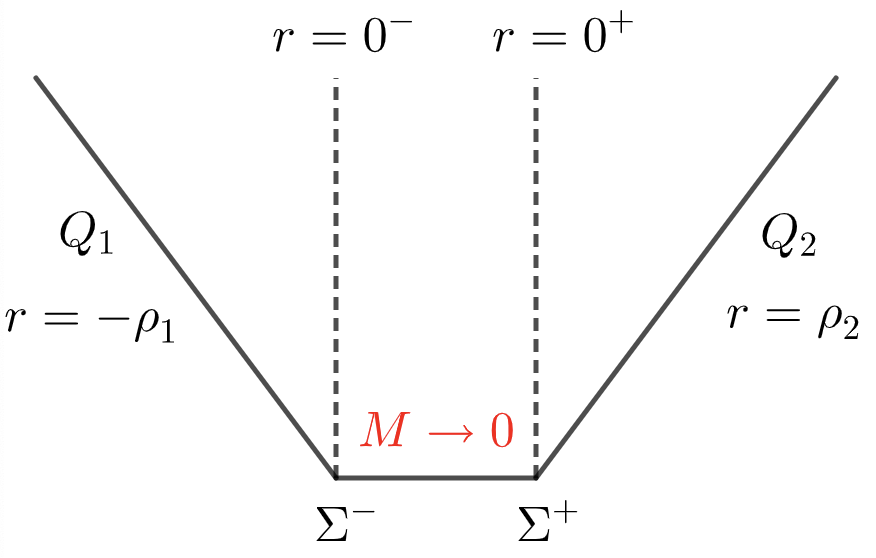}
\caption{Wedge holography from AdS/BCFT. Here BCFT lives on manifold $M$ with two boundaries $\Sigma^-$ and $\Sigma^+$. The dotted lines denote the zero-tension branes. The left bulk region $(-\rho_1\le r\le 0^-)$ is dual to the edge mode of BCFT on the left boundary $\Sigma^-$. And similar for the right bulk region $(0^+\le r\le \rho_2)$. By taking the limit $M\to 0$,  we obtain wedge holography from AdS/BCFT.}
\label{WedgefromAdSBCFT}
\end{figure}

\rrr{Wedge holography meets DGP gravity}
The action of wedge holography with DGP gravity (DGP wedge holography) on the brane is given by 
\begin{eqnarray}\label{action}
I&=&\int_W dx^{d+1}\sqrt{|g|}\Big(R+d(d-1)\Big) \nonumber\\
&&+2\int_{Q} dx^d\sqrt{|h_Q|}(K-T_a+\lambda_a R_{Q}),
\end{eqnarray}
where we have set $16\pi G_N=1$ and AdS radius $L=1$,
$W$ is the wedge space in bulk, $Q=Q_1\cup Q_2$ denote two end-of-the-world branes, $K$ is the extrinsic curvature, $R_{Q}$ is the DGP term (Ricci scalar) on the brane, $T_a$ and $\lambda_a$ with $a=1,2$ are parameters. See Fig.\ref{Wedge} and Fig.\ref{WedgefromAdSBCFT} for the geometry.  As shown in Fig.\ref{WedgefromAdSBCFT}, wedge holography can be derived from the zero-volume limit $M\to 0$ of AdS/BCFT \cite{Akal:2020wfl}. Here dotted lines denote the zero-tension branes. The left bulk region $(-\rho_1\le r\le 0^-)$ is dual to the edge mode of BCFT on the left boundary $\Sigma^-$. And similar for the right bulk region. Strong support for this partial duality comes from
boundary entropy and boundary central charges \cite{Takayanagi:2011zk,Miao:2020oey}. To have massless gravity on the brane \cite{Hu:2022lxl}, we impose Neumann boundary condition (NBC) on both branes 
\begin{eqnarray}\label{NBC}
K^{ij}-(K-T_a+\lambda_a R_{Q}) h_Q^{ij}+2 \lambda_a R^{ij}_Q=0.
\end{eqnarray}
There is an exact solution
\cite{Miao:2020oey}
\begin{eqnarray}\label{metric}
ds^2=dr^2+\cosh^2(r) h_{ij}(y) dy^i dy^j, \ -\rho_1 \le r\le \rho_2,
\end{eqnarray}
provided that $h_{ij}$ is the metric of Einstein manifold $R_{h\ ij}=-(d-1)h_{ij}$ and  the parameters obey $T_a=(d-1) \tanh(\rho_a)-\lambda_a \frac{(d-1)(d-2)}{\cosh^2(\rho_a)}.$
 Substituting (\ref{metric}) into the action (\ref{action}) and integrating $r$, we get an effective action on each brane
  \begin{eqnarray}\label{effective action}
I_a=\frac{1}{16\pi G^a_{\text{eff N}}}\int_{Q_a} \sqrt{|h|} \Big( R_h+(d-1)(d-2) \Big),
 \end{eqnarray} 
where $G^a_{\text{eff N}}$ denotes the effective Newton's constant
  \begin{eqnarray}\label{effective Newton constant}
\frac{1}{16\pi G^a_{\text{eff N}}}=\int_0^{\rho_a} \cosh^{d-2}(r) dr+2\lambda_a \cosh^{d-2}(\rho_a).
 \end{eqnarray} 
Let us make some comments. 
{\bf 1}. For the special class of solution (\ref{metric}), the induced metric $h_{ij}$ on the two spatially-separated branes constantly fluctuates together. It is a typical characteristic of standing waves and cannot spread signals instantaneously. Recall the standing waves between two parallel plates in the real world, i.e., $\phi(x)=\cos( n \pi x /L) e^{-i \omega t}$. The oscillations on the two plates are the same for even integer $n$ ($\phi(0)=\phi(L)$) and are opposite for odd integer $n$ ($\phi(0)=-\phi(L)$). Everything goes well for the standing wave in the real world. Our case (\ref{metric}) is the same: just the zero modes of the standing wave in a wedge space. In fact, if one instantly changes the metric on the left brane, the metric on the right brane will not be affected until the gravitational waves propagate from bulk.
 {\bf 2}. As shown in (\ref{effective action}), there is Einstein gravity (zero modes) on the brane. There are also massive modes \cite{Miao:2023unv}, which generally make the metric fluctuations on the two branes different. The key point is that, unlike Karch-Randall braneworld \cite{Karch:2000ct}, wedge holography includes massless gravitons on the branes. At low energy, the massive modes decouple and only the massless mode can be excited. As a result, Einstein gravity behaves as a low-energy effective theory on the brane. {\bf 3}. Inspired by \cite{Takayanagi:2011zk, Akal:2020wfl}, we divide the bulk into two parts by a zero-tension brane at $r=0$. See Fig. \ref{WedgefromAdSBCFT}. Then, we take the region $-\rho_1\le r\le 0$ to calculate Newton's constant on $Q_1$. This partial region is dual to the edge mode in AdS/BCFT. {\bf 4}. We require $G^a_{\text{eff N}} >0$ so that the dual CFT has positive central charges \cite{Miao:2020oey}. 
 {\bf 5}. Following \cite{Akal:2020wfl,Miao:2020oey}, one can check that DGP wedge holography yields the correct universal terms of entanglement/R\'enyi entropy, Weyl anomaly, two point functions of stress tensor. This is a strong support for DGP wedge holography. {\bf 6}. Imposing null energy condition on the brane, one can prove the holographic c-theorem. See the appendix for details. 

\rrr{Massless island}
Let us discuss the Page curve for the eternal two-side black hole, which is dual to the thermofield double state of CFTs \cite{Maldacena:2001kr}. In the usual double holography, the black hole in the AdS brane is coupled to an asymptotic non-gravitational bath on the AdS boundary. In this paper, we consider gravitational bath in wedge holography.
Let us first study the island phase (blue curve of Fig.\ref{Wedge}). For wedge holography without DGP terms, \cite{Geng:2020fxl} finds that the island region disappears outside the horizon. 
Below, we show how to recover islands by employing negative DGP gravity. 

Let us focus on the black string in bulk
\begin{eqnarray}\label{BHmetric}
ds^2=dr^2+\cosh^2(r) \frac{\frac{dz^2}{f(z)}-f(z)dt^2+dy_{\hat{i}}^2}{z^2},
\end{eqnarray}
where there is an AdS-Schwarzschild black hole with $f(z)=1-z^{d-1}$ on each brane, $\hat{i}$ 
 runs from 1 to $d-2$.  One chooses the black hole on the weak-gravity brane as the bath approximately \cite{Geng:2020fxl}. From (\ref{action}), 
 \cite{Chen:2020uac} derives the holographic entanglement entropy for wedge holography with DGP terms
\begin{eqnarray}\label{HEE}
S={\text{min}}\Big(  4\pi \int_{\Gamma} dx^{d-1} \sqrt{\gamma}+8\pi \lambda_a \int_{\partial \Gamma} dx^{d-2} \sqrt{\sigma} \Big),
\end{eqnarray}
where $\Gamma$ labels RT surface, $\partial \Gamma=\Gamma\cap Q$ is the intersection of the RT surface and branes. 
\cite{Chen:2020uac} also provides non-trivial tests for the entropy formula (\ref{HEE}). 
Substituting the embedding functions $t=\text{constant}, z=z(r)$ into (\ref{BHmetric}) and (\ref{HEE}), we get the area functional $S/(4\pi)$
\begin{eqnarray}\label{areaisland}
A_{\text{I}}&=&V\int_{-\rho_1}^{\rho_2} dr\frac{\cosh^{d-2}(r)}{z^{d-2}} \sqrt{1+\frac{\cosh^{2}(r) z'^2}{z^{2}f(z)}}\nonumber\\
&&+V\sum_{a=1}^2\frac{2\lambda_a \cosh^{d-2}(\rho_a)}{z^{d-2}_a},
\end{eqnarray}
where $\text{I}$ denotes the island phase, $V=\int dy^{d-2}$ is the horizontal volume, $z'=\partial_r z(r)$ and $z_a=z[(-)^a \rho_a]$ are boundary values of $z$ on the two branes.

Let us first discuss the case without DGP terms, i.e., $\lambda_a=0$. Then, by using $0\le z\le 1$ and $f(z)\ge 0$ outside the horizon, we get $A_{\text{I}}\ge V\int_{-\rho_1}^{\rho_2} dr\cosh^{d-2}(r)=A_{\text{BH}}$, which means that the black hole horizon has the minimal area $A_{\text{BH}}$ and is the RT surface in the island phase. Another way to see this is as follows. 
Without DGP terms, the RT surface must end orthogonally on both branes. Otherwise, it cannot be a minimal area surface. This orthogonal condition is so strong that no extremal surfaces except the horizon can satisfy \cite{Geng:2020fxl}. 
When the RT surface (blue curve of Fig. \ref{Wedge}) coincides with the horizon (dotted line of Fig. \ref{Wedge}), 
both the radiation and island regions vanish, yielding zero entanglement entropy of radiation \cite{Geng:2020fxl,Geng:2023qwm}. We stress that the entanglement island and radiation region disappear at the same time. Naturally, a zero radiation region has no information loss and, thus, no need for an entanglement island. However, physically, we are more interested in a non-trivial radiation region. 

Let us go on to discuss the effects of DGP terms. The bulk term of (\ref{areaisland}) decreases with $z$ and becomes minimal on the horizon $z=1$ \cite{Geng:2020fxl}, while the boundary term of (\ref{areaisland}) increases with $z_a$ for negative $\lambda_a$. 
Due to the competition of these two terms, the area (\ref{areaisland}) could minimize outside the horizon if at least one of $\lambda_a$ is negative. There is another way to understand this. Note that the RT surface is no longer perpendicular to the brane when there is a DGP term \cite{Chen:2020uac}. As a result, the no-go theorem of \cite{Geng:2020fxl} becomes invalid. Let us explain more. For a well-defined action principle, one must impose suitable boundary conditions so that the variation of (\ref{areaisland}) vanishes on the boundary. We can impose either Dirichlet boundary condition (DBC) $\delta z_a=0$ or NBC
\begin{eqnarray}\label{NBCisland}
\frac{z_a'}{f(z_a)\sqrt{1+\frac{\cosh^{2}(\rho_a) z_a'^2}{z_a^{2}f(z_a)}}}=\frac{2\lambda_a(-1)^a(d-2)z_a}{\cosh^2(\rho_a)},
\end{eqnarray}
where the repeated index `a' does not sum. In the island phase, we choose NBC (\ref{NBCisland}) since it can give a smaller area than DBC by changing the endpoint $z_a$ on the branes. Note that $z_a'=0$ means that the RT surface is perpendicular to the brane. From (\ref{NBCisland}), it is clear that $z'_a$ is non-zero due to the DGP parameter $\lambda_a$. This avoids the strong orthogonal condition of \cite{Geng:2020fxl} and allows a non-trivial island outside the horizon. To see this clearly, let us turn the logic around. We can solve a class of extremal surfaces outside the horizon using the Euler-Langrangian equation derived from (\ref{areaisland}). For any such extremal surface, we can determine $z_a, z_a'$ on the branes $r_a=(-)^a\rho_a$ and then derive $\lambda_a$ from NBC (\ref{NBCisland}). Now return to our problem. For the parameters 
fixed above, the RT surface with the minimal area (\ref{areaisland}) is just the input extremal surface outside the horizon, which satisfies both the Euler-Langrangian equation and NBC. By choosing suitable parameters, 
we can ensure the input extremal surface is minimal instead of maximal \cite{Miao:2023unv}. Thus the massless island indeed exists in wedge holography with suitable DGP terms.    
Without loss of generality, we take $(\rho_1=0.5, \rho_2=1.1, \lambda_1=0, \lambda_2\approx -0.243)$ and $d=4, V=1$ as an example, which gives
\begin{eqnarray}\label{G1G2}
0<G^1_{\text{eff N}} \approx 0.037 <  G^2_{\text{eff N}}\approx 0.064. 
\end{eqnarray}
Since the left brane has a smaller effective Newton's constant, it corresponds to the weak gravity region and can be taken as the gravitational bath. By minimizing the area functional (\ref{areaisland}), we determine numerically the RT surface, which has a smaller area than that of black holes
\begin{eqnarray}\label{SISBH}
A_{\text{I}}\approx 0.835 < A_{\text{BH}} \approx 0.853,
\end{eqnarray}
where we focus on half of the two-side black hole. This RT surface (also called island surface) starts at $z_1\approx 0.970$ on the left brane and ends at $z_2\approx 0.625$ on the right brane. So the radiation region (red line of Fig.\ref{Wedge}) locates at $z\ge z_1\approx0.970$ on the left brane.  

\begin{figure}[t]
\centering
\includegraphics[width=8.2cm]{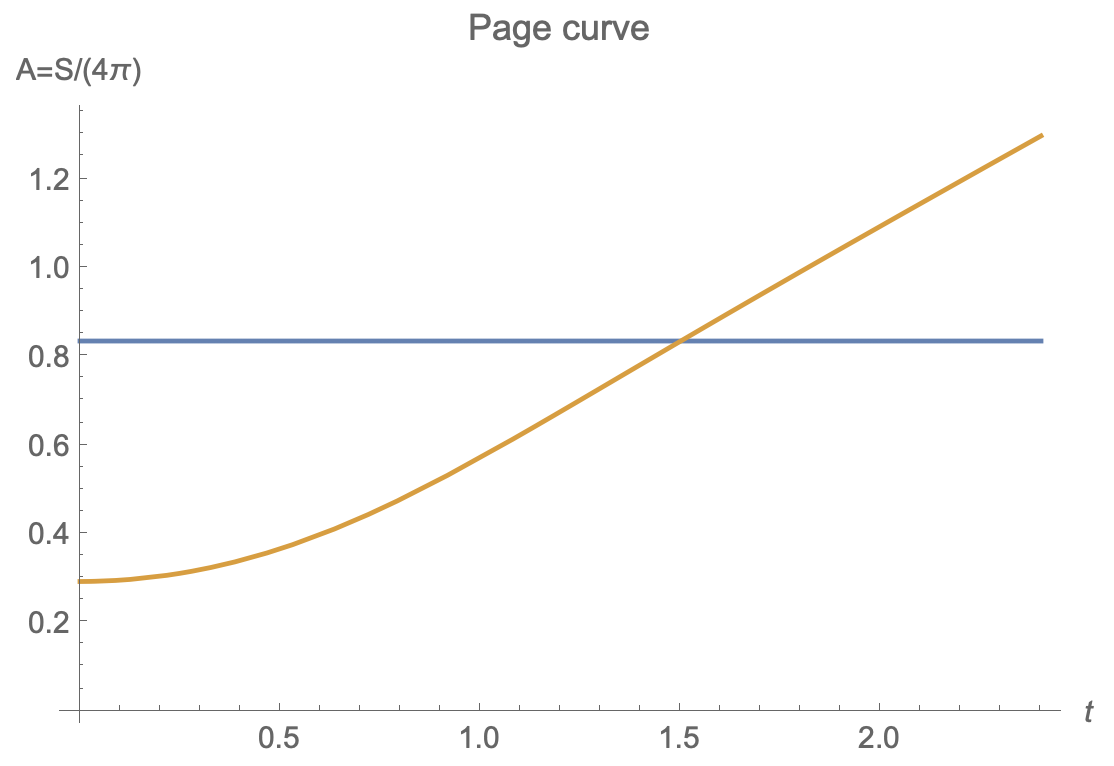}
\caption{Page curve for DGP wedge holography. The orange and blue curves denote the area of HM surface and island surface, respectively. The entanglement entropy firstly increases with time (orange line) and then becomes a constant (blue line), which recovers the Page curve of eternal black holes.}
\label{Pagecurve}
\end{figure}

\rrr{Universality of no-island phase}
Let us go on to study the RT surface in the no-island phase, also called Hartman-Maldacena (HM) surface (orange curve of Fig.\ref{Wedge}). The HM surface starts at $z=z_1$ on the left bath brane, ends on the horizon at the start $t=0$, and then passes the horizon at $t>0$. Let us first study the case at $t=0$. By adjusting the endpoint on the horizon, we derive the HM surface with minimal area
\begin{eqnarray}\label{Snoisland}
A_{\text{N}}\approx 0.293, \ \text{at t=0}
\end{eqnarray}
which is smaller than $A_{\text{I}}$ (\ref{SISBH}). Here $\text{N}$ denotes the no-island phase. Thus the no-island phase dominates at the beginning. At $t>0$, the HM surface passes through the horizon, and it is convenient to use the infalling Eddington-Finkelstein coordinate $dv=dt-\frac{dz}{f(z)}$. Assuming the embedding function $v=v(z), r=r(z)$, we obtain the area functional 
\begin{eqnarray}\label{areanoisland}
A_{\text{N}}&=&\frac{S}{4\pi}=V\int_{z_1}^{z_{\text{max}}} dz\frac{\cosh^{d-2}(r)}{z^{d-2}} \nonumber\\
&&\ \ \ \ \times \sqrt{r'^2-\frac{\cosh^{2}(r)}{z^{2}}v'(2+f(z)v')) }
\end{eqnarray}
and the time of CFTs on the left bath brane $t_1=t(z_1)=-\int_{z_1}^{z_{\text{max}}} \Big( v'(z)+\frac{1}{f(z)}\Big) dz$,
where $z_{\text{max}}\ge 1$ is the turning point of the two-side black hole, which obeys $v'(z_{\text{max}})=-\infty$ and $t(z_{\text{max}})=0$ \cite{Carmi:2017jqz}.
Note that we have set $\lambda_1=0$ as in the above example.
For simplicity, we label $t_1$ by $t$ in this paper. By numerical analysis, we find that $r$ approaches zero for large times. Thus we have $\lim_{t\to \infty} A_{\text{N}}=V\int_{z_1}^{z_{\text{max}}}\frac{ dz}{z^{d-2}} \sqrt{-\frac{v'(2+f(z)v'))}{z^{2}} }$,
which is the same as the volume conjecture of holographic complexity \cite{Susskind:2014rva,Stanford:2014jda} for AdS-Schwarzschild black hole in $d$ spacetime dimensions. 
This raises the question if there is a deep relation between entanglement entropy and complexity. 
Following \cite{Carmi:2017jqz}, we get
\begin{eqnarray}\label{entropylargetime}
\lim_{t\to \infty} \frac{d A_{\text{N}}}{dt}=\frac{V}{2},
\end{eqnarray}
which shows that area of HM surface increases linearly over time at late times. 
Note that this late-time behavior is independent of the parameters $(\rho_a, \lambda_a)$
as long as the radiation region is non-zero. The HM surface vanishes for a zero radiation region, and the entanglement entropy of radiation remains zero during the time evolution \cite{Geng:2023qwm}.
By numerical calculations, we derive the general time dependence of $A_{\text{N}}$ and draw the Page curve in Fig. \ref{Pagecurve},
where $A$ and $t$ are half that of a two-side black hole.
At early times, the no-island phase (orange line) dominates, and the entanglement entropy increases with time. After the Page time, the island phase (blue line) dominates, and the entanglement entropy becomes a constant. So the Page curve of eternal black hole is recovered.

Some comments are in order. {\bf 1}. As discussed above, to have a nontrivial island outside the horizon, at least one of $\lambda_a$ is negative. The negative DGP gravity, however, leads to a massive ghost \cite{Miao:2023mui}. We will discuss briefly the ghost problem below. {\bf 2}. The ghost problem can be resolved if one imposes DBC instead of NBC on the brane (like one imposes DBC on the AdS boundary in AdS/BCFT with a non-gravitational bath). Although DBC fixes the induced metric on the brane, it allows dynamical gravity from the fluctuations of extrinsic curvatures \cite{Chu:2021mvq}. For DBC, the negative DGP term plays the role of counter terms, as in AdS/CFT.  
One can check the ghost mode disappears in the mass spectrum with DBC. Unfortunately, although DBC removes the massive ghost, it also removes the massless graviton. See discussions below. Thus, one cannot simultaneously have the island and massless graviton in a ghost-free braneworld model.  {\bf 3}. The entanglement entropy in this paper is renormalized and finite since the branes are located at finite places instead of asymptotic infinity. {\bf 4}. Due to the similarity to wedge holography, the massless island also exists in cone holography with negative DGP gravity \cite{Miao:2021ual, Li:2023fly}, which suffers the same ghost problem.   
 
\rrr{The ghost problem}
Now, we show the spectrum of negative DGP gravity includes a massive ghost. For simplicity, we focus on the parameters ($\lambda_1=0, \lambda_2<0$), which is the case discussed above. Consider the perturbation around the background metric (\ref{metric})
 \begin{eqnarray}\label{perturbationmetric}
ds^2=dr^2+\cosh^2(r)\left(h_{ij}(y) +H(r) h^{(1)}_{ij}(y) \right)dy^idy^j,\nonumber\\
\end{eqnarray}
where $h^{(1)}_{ij}$ denotes the metric perturbation, and the brane locations are unchanged at the linear perturbation. Imposing the transverse traceless gauge for $h^{(1)}_{ij}(y)$ and separating variables of Einstein equations, we get \cite{Miao:2023mui}
\begin{eqnarray}
\cosh^2(r) H''(r)+\frac{d}{2} \sinh(2r)H'(r)+m^2 H(r)=0, \label{EOMMBCmassiveH}
\end{eqnarray}
where $m$ denotes the mass of gravitons on the brane. Solving (\ref{EOMMBCmassiveH}), we obtain
 \begin{eqnarray}\label{massiveHsolution}
H_m(r)=\text{sech}^{\frac{d}{2}}(r) \left(c_1 P_{l_m}^{\frac{d}{2}}(x)+c_2 Q_{l_m}^{\frac{d}{2}}(x)\right),
\end{eqnarray} 
where $x=-\tanh r$, $P_{l_m}^{\frac{d}{2}}(x)$ and $ Q_{l_m}^{\frac{d}{2}}(x)$ are the Legendre polynomials, $c_1$ and $c_2$ are integral constants and $2l_m=\sqrt{(d-1)^2+4 m^2}-1$. For the ansatz (\ref{perturbationmetric}), the NBC (\ref{NBC}) with $\lambda_1=0$ becomes 
\begin{eqnarray} \label{NBCperturbation}
H_m'(-\rho_1)=0, \ H_m'(\rho_2)=2 \lambda_2 m^2 H_m(\rho_2) \text{sech}^2(\rho_2).
\end{eqnarray}
The above boundary conditions fix the mass spectrum and the ratio $c_1/c_2$ for each Kaluza-Klein (KK) mode. 
The orthogonal relation of KK modes implies the inner product \cite{Miao:2023mui} 
 \begin{eqnarray}\label{orthogonal-gravity}
\langle H_m, H_m\rangle&=&\int_{-\rho_1}^{\rho_2}\cosh(r)^{d-2}H^2_m(r)dr\nonumber\\
&&+2 \lambda_2 \cosh^{d-2}(\rho_2) H^2_m(\rho_2),
\end{eqnarray}
up to a normalization of $H_m(r)$. 
We remark that the inner product (\ref{orthogonal-gravity}) becomes the sum of the reciprocal of Newton's constant (\ref{effective Newton constant}) for the massless mode with $ m^2=0, H_m(r)=1$. For negative $\lambda_2$, the inner product (\ref{orthogonal-gravity}) could become negative for some modes. Take the parameters $(\rho_1=0.5, \rho_2=1.1, \lambda_1=0, \lambda_2\approx -0.243, d=4)$ as an example, we find a ghost with negative inner product
 \begin{eqnarray}\label{orthogonal-gravity1}
\langle H_m, H_m\rangle\approx -0.369, \ \text{for} \ m^2\approx -2.780,
\end{eqnarray}
where we have chosen the normalization $H_m(\rho_2)=1$. The above ghost is also a tachyon with negative $m^2$. One can prove there is always a ghost mode for $\lambda_1=0, \lambda_2<0$. The fastest way to see this is by applying the spectrum identity \cite{Miao:2023mui}
\begin{eqnarray}\label{nicerelationDGP}
\sum_{m} \frac{H_m^2(\rho_2)}{\langle H_m, H_m \rangle} = \frac{1}{2\lambda_2}.
\end{eqnarray} 
To satisfy (\ref{nicerelationDGP}) for negative $\lambda_2$, at least one inner product of KK mode should be negative. Note that the spectrum identity becomes divergent in the limit $\lambda_2\to 0$, implying a phase transition of the spectrum. That is reasonable because DGP gravity brings an additional mode on the brane, which changes the spectrum discontinuously.  

Now let us discuss what happens if we impose Dirichlet boundary condition (DBC) on the second brane $Q_2$
\begin{eqnarray} \label{DBCperturbation}
H_m(\rho_2)=0.
\end{eqnarray}
One can impose either NBC $H_m'(-\rho_1)=0$ or DBC $H_m(-\rho_1)=0$ on the first brane $Q_1$. 
For the above boundary conditions, the inner product becomes 
 \begin{eqnarray}\label{orthogonal-gravity-DBC}
\langle H_m, H_m\rangle_{\text{DBC}}=\int_{-\rho_1}^{\rho_2}\cosh(r)^{d-2}H^2_m(r)dr >0,
\end{eqnarray}
which is always positive, suggesting no ghost modes for DBC. Unfortunately, DBC removes the massless mode, too. From EOM (\ref{EOMMBCmassiveH}), we solve $H_m(r)=1$ for the massless mode with $m^2=0$. Clearly, $H_m(r)=1$ disobeys the DBC (\ref{DBCperturbation}).

\rrr{Conclusion and open question} This paper shows that entanglement islands can exist in long-range gravity in a braneworld model with negative DGP gravity. However, negative DGP gravity suffers the ghost problem, which is not well-defined. On the other hand, the positive DGP gravity is ghost-free but has no entanglement island. Our results support the view there is no entanglement island in a well-defined theory of massless gravity if one divides the radiation and black hole regions by minimizing entanglement entropy. According to \cite{Geng:2023qwm}, one always gets a zero radiation region with zero entanglement entropy in such a partition. As expected, a zero radiation region results in a vanishing island. However, we are more interested in a non-zero radiation region. It raises the question of whether other non-trivial partitions of the radiation region exist. Naturally, one can consider the radiation region observed by an observer \cite{Witten:2023qsv}, which is generally non-zero. Whether this scheme works is an open question and deserves further study.


\rrr{Aacknowledgments} 
We thank T. Takayanagi, X. Dong, Y. Pang, H.J. Wang and D.q. Li for valuable comments. This work is supported by the National Natural Science Foundation of China (No.12275366 and No.11905297).

\appendix
\section{Appendix: holographic c-theorem}
Let us proves the holographic c-theorem on the brane with DGP terms, which provides another support for DGP wedge holography. Note that the holographic c-theorem on the brane is actually the holographic g-theorem in AdS/BCFT \cite{Takayanagi:2011zk}. Following \cite{Takayanagi:2011zk}, we focus on an AdS spacetime in bulk.  Performing coordinate transformations $\hat{z}=z /\cosh(r), x=z \tanh(r)$, we rewrite AdS metric (\ref{BHmetric}) with $f(z)=1$ into the Poincare form
\begin{eqnarray}\label{AdSmetric}
ds^2=\frac{d\hat{z}^2-dt^2+dx^2+dy_{\hat{i}}^2 }{\hat{z}^2},
\end{eqnarray}
where matter fields live on the branes $x=(-1)^aF(\hat{z})$. Then the NBC (\ref{NBC}) becomes
\begin{eqnarray}\label{NBCmatter}
K^{ij}-(K-T_a+\lambda_a R_{Q}) h_Q^{ij}+2 \lambda_a R^{ij}_Q=\frac{1}{2}T^{ij}_{\text{M}},
\end{eqnarray}
with $T^{ij}_{\text{M}}$ the matter stress tensor. We impose the null energy condition $T^{ij}_{\text{M}} N_i N_j\ge 0$ on the brane,
where $N^i=(\frac{1}{\sqrt{1+(F'(z))^2}},\pm 1,0)$. It yields
\begin{eqnarray}\label{NEN1}
\frac{\left(\sqrt{1+F'(\hat{z})^2}+2(d-2) \lambda_aF'(\hat{z}) \right)F''(\hat{z}) }{ \left(1+F'(\hat{z})^2\right)^{2}}\le 0.
\end{eqnarray}
Multiplying the LHS of (\ref{NEN1}) by a positive factor $\left(1+F'(\hat{z})^2\right)^{\frac{d}{2}}$ and integrating along $z$, we get the c-function 
\begin{eqnarray}\label{cfunction}
c(\hat{z})&=&2\lambda_a\left(1+F'(\hat{z})^2\right)^{\frac{d}{2}-1}\nonumber\\
&&+\, _2F_1\left(\frac{1}{2},\frac{3-d}{2};\frac{3}{2};-F'(\hat{z})^2\right)F'(\hat{z}),
\end{eqnarray}
where we have added an integral constant $2\lambda_a$. By construction, $c(\hat{z})$ obeys the c-theorem $c'(\hat{z}) \le 0$, where $1/\hat{z}$ denotes the energy scale. At the UV fixed point, we have $\lim_{\hat{z}\to 0} F'(\hat{z})=\sinh(\rho_a)$ \cite{Takayanagi:2011zk}, thus we get
\begin{eqnarray}\label{cfunction1}
&&\lim_{\hat{z}\to 0}c(\hat{z})=\frac{1}{16\pi G^a_{\text{eff N}}}=2 \lambda_a \cosh ^{d-2}(\rho_a) \nonumber\\
&&\ \ \ \ +\, _2F_1\left(\frac{1}{2},\frac{3-d}{2};\frac{3}{2};-\sinh^2(\rho_a )\right)\sinh (\rho_a ),
\end{eqnarray}
which is equal to the inverse of the effective Newton's constant (\ref{effective Newton constant}). In the above derivations, we have use the formula $\int \cosh^{d-2}(r) dr=\int (1+x^2)^{\frac{d-3}{2}} dx=x \, _2F_1\left(\frac{1}{2},\frac{3-d}{2};\frac{3}{2};-x^2\right)$, where $x=\sinh(r)$. Recall that the A-type central charge of dual CFTs is proportional to $1/G^a_{\text{eff N}}$, so the c-function (\ref{cfunction}) reduces to A-type central charge at the UV fixed point. Now we finish the proof of holographic c-theorem on the brane with DGP gravity. Note that $\lambda_a$ is arbitrary in the above proof. We further require that the c-function (\ref{cfunction}) is positive at least at one point of $F'(\hat{z})$. Equivalently, the effective Newton's constant (\ref{cfunction1}) is positive at least for one $\rho_a$. It leads to a lower bound of $\lambda_a$. For instance, we have $\lambda_a\gtrsim -0.300,\ \text{for d=4}.$


\end{document}